\documentclass[review]{elsarticle}
\usepackage[utf8]{inputenc}				
\usepackage{hyperref,amsmath,amssymb,amsbsy,bbold,color,graphics,graphicx,bigints,relsize,cancel}
\usepackage{lineno,hyperref,appendix}

\modulolinenumbers[5]
\pdfoutput=1
\allowdisplaybreaks

\journal{Journal of \LaTeX\ Templates}

\begin{document}

\begin{frontmatter}

\title{Onsager-Casimir reciprocal relations}

\author{Sylvain D. Brechet\fnref{myfootnote}}
\address{Institute of Physics, Station 3, Ecole Polytechnique F\'ed\'erale de Lausanne\,-\,EPFL, CH-1015 Lausanne,\,Switzerland}

\ead{sylvain.brechet@epfl.ch}
	
\begin{abstract}

The Onsager reciprocal relations are established within the phenomenological framework of the thermodynamics of irreversible processes. In order to do so, the dissipated power densities associated to scalar and vectorial processes are written as positive semi-definite quadratic forms of the corresponding generalised forces, as required by the local expression of the second law in the neighbourhood of the equilibrium. The antisymmetric part of the scalar and vectorial Onsager matrices do not contribute to the dissipation, which yields the scalar and vectorial Onsager reciprocal relations. Furthermore, the positive semi-definite quadratic forms of the generalised scalar and vectorial forces are invariant under time reversal, which yields the scalar and vectorial Casimir-Onsager reciprocal relations, that are a generalisation of the Onsager reciprocal relations.

\end{abstract}

\end{frontmatter}


\tableofcontents
	

\section{Introduction}
\label{Introduction}
	
\noindent In 1931, Lars Onsager determined the symmetries of the phenomenological coefficients of the linear applications mapping the generalised forces to the generalised forces in thermodynamic systems that are locally at equilibrium but globally out of equilibrium.~\cite{Onsager:1931a,Onsager:1931b} These relations, known as the Onsager reciprocal relations, were established within the framework of statistical physics. The approach taken by Onsager is quite general in the sense that he considered a mechanical system consisting of particles. His theoretical work is based on the microscopic reversibility, which requires the symmetry of the equilibrium fluctuation correlations. It widely believed that these relations cannot be derived within the phenomenological framework of the thermodynamics of irreversible processes.~\cite{Brechet:2019,Kubo:1985,DeGroot:1985,Marder:2010} An attempt was made by Christian Gruber to show that such an assumption is unwarranted for the very particular case of a system of particles characterised by a global entropy.~\cite{Gruber:1977} In this publication, we will show that for a thermodynamic system that is locally at equilibrium but globally out of equilibrium~\cite{Brechet:2019}, the Onsager reciprocal relations, whether of scalar or vectorial nature, are a direct consequence of the local expression of the second law of thermodynamics.

In 1945, Hendrick Casimir showed that the Onsager reciprocal relations do no longer hold in general if different generalised forces have not the same symmetry under time reversal.~\cite{Casimir:1945} He showed in particular that in the presence of a magnetic field that changes sign under time reversal these relations do no longer hold.~\cite{Callen:1985} The generalisation of the reciprocal relations that take into account the symmetries under time reversal are known as the Onsager-Casimir reciprocal relations. In this article, we will also show how these relations can be derived in a purely phenomenological approach by taking into account the symmetry under time reversal in addition to the local expression of the second law of thermodynamics.

This publication is structured as follows : in Sec.~\ref{Internal power density for irreversible processes}, we review the mathematical structure of the scalar and vectorial internal power densities of irreversible processes in a continuous medium consisting of electrically charged chemical substances undergoing coupled chemical reactions. The Onsager reciprocal relations between the scalar generalised forces and currents densities are established in Sec.~\ref{Scalar Onsager reciprocal relations}. The scalar Onsager-Casimir reciprocal relations are derived in Sec.~\ref{Scalar Onsager-Casimir reciprocal relations}. In Sec.~\ref{Scalar Onsager reciprocal relations}, we establish the Onsager reciprocal relations between the vectorial generalised forces and current densities. The vectorial Onsager-Casimir reciprocal relations are derived in Sec.~\ref{Scalar Onsager-Casimir reciprocal relations}. Finally, we conclude our analysis of the Onsager-Casimir relations in Sec.~\ref{Conclusion}.


\section{Internal power density for irreversible processes}
\label{Internal power density for irreversible processes}

\noindent In order to establish the Onsager relations entirely within the framework of the thermodynamics of irreversible processes, we consider a continuous medium consisting of $r$ electrically charged chemical substances, undergoing $n$ coupled chemical reactions in the absence of shear and vorticity. The irreversible thermodynamic evolution of the system is described by the internal power density $p^{\,\text{int}}$ accounting for the dissipation, 
\begin{align}\label{power density int}
p^{\,\text{int}} = \sum_{a=1}^{n}\,\omega_a\,\mathcal{A}_a + \tau^{\,\text{fr}}\,\left(\boldsymbol{\nabla}\cdot\boldsymbol{v}\right) + \boldsymbol{j}_s\cdot\left(-\,\boldsymbol{\nabla}\,T\right) + \sum_{A=1}^{r}\,\boldsymbol{j}_A\cdot\big(-\,\boldsymbol{\nabla}\,\bar{\mu}_A\big) \geqslant 0
\end{align}
where $\omega_a$ is the rate density and $\mathcal{A}_a$ is the affinity of the chemical reaction $a$, $\tau^{\,\text{fr}}$ is the isotropic internal friction and $\left(\boldsymbol{\nabla}\cdot\boldsymbol{v}\right)$ is the local volume expansion rate density, $\boldsymbol{j}_s$ is the entropy current density and $\boldsymbol{\nabla}\,T$ is the temperature gradient, $\boldsymbol{j}_A$ is the matter current density and $\bar{\mu}_A$ is the electrochemical potential of substance $A$. The first two power density terms are the product of scalar quantities describing the dissipation due to chemical reactions and to the internal friction within the continuous medium. They will be referred to as the scalar internal power density $p^{\,\text{int}}_S$. The last two power density terms are the product of vector quantities describing the dissipation due to the transport of chemical substances. They will be referred to as the vectorial internal power density $p^{\,\text{int}}_V$. Thus,
\begin{equation}\label{power density int S and V}
p^{\,\text{int}} = p^{\,\text{int}}_S + p^{\,\text{int}}_V \geqslant 0
\end{equation}
Since scalar and vectorial terms have different symmetries, according to the Curie principle, the scalar internal power density $p^{\,\text{int}}_S$ and the vectorial internal power density $p^{\,\text{int}}_V$ have to be separately irreversible,
\begin{equation}\label{power density int S and V bis}
p^{\,\text{int}}_S \geqslant 0 \qquad\text{and}\qquad p^{\,\text{int}}_V \geqslant 0
\end{equation}
Following Onsager's approach, the scalar internal power density~\eqref{power density int S and V bis} can be formally written in matrix form as the product of a line vector and a column vector consisting of $n+1$ scalar components,
\begin{equation}\label{power density int S}
p^{\,\text{int}}_S = 
\left(\boldsymbol{\nabla}\cdot\boldsymbol{v},\mathcal{A}_1,\ldots,\mathcal{A}_n\right)
\begin{pmatrix} 
\tau^{\,\text{fr}}\\
\omega_1\\
\vdots\\
\omega_n
\end{pmatrix} \geqslant 0
\end{equation}
Similarly, the vector internal power density~\eqref{power density int S and V} can be formally written as a matrix product of a line vector and a column vector in consisting of $r+1$ vector components,
\begin{equation}\label{power density int V}
p^{\,\text{int}}_V = 
\left(-\,\boldsymbol{\nabla}\,T,-\,\boldsymbol{\nabla}\,\bar{\mu}_1,\ldots,-\,\boldsymbol{\nabla}\,\bar{\mu}_r\right)
\cdot\begin{pmatrix} 
\boldsymbol{j}_s\\
\boldsymbol{j}_1\\
\vdots\\
\boldsymbol{j}_r
\end{pmatrix} \geqslant 0
\end{equation}
The generalised scalar force column vector $F \in \mathbb{R}^{n+1}$ and the generalised vectorial force column vector $\boldsymbol{F} \in \mathbb{R}^{n+1}$ are defined as,
\begin{equation}\label{generalised force vectors}
F =
\begin{pmatrix} 
\boldsymbol{\nabla}\cdot\boldsymbol{v}\\
\mathcal{A}_1\\
\vdots\\
\mathcal{A}_n
\end{pmatrix}
\qquad\text{and}\qquad
\boldsymbol{F} = 
\begin{pmatrix} 
-\,\boldsymbol{\nabla}\,T\\
-\,\boldsymbol{\nabla}\,\bar{\mu}_1\\
\vdots\\
-\,\boldsymbol{\nabla}\,\bar{\mu}_r
\end{pmatrix}
\end{equation}
The generalised scalar current density column vector $j \in \left(\mathbb{R}^3\right)^{r+1}$ and the generalised vectorial current density column vector $\boldsymbol{j} \in \left(\mathbb{R}^3\right)^{r+1}$ are defined as,
\begin{equation}\label{generalised current density vectors}
j =
\begin{pmatrix} 
\tau^{\,\text{fr}}\\
\omega_1\\
\vdots\\
\omega_n
\end{pmatrix}
\qquad\text{and}\qquad
\boldsymbol{j} = 
\begin{pmatrix} 
\boldsymbol{j}_s\\
\boldsymbol{j}_1\\
\vdots\\
\boldsymbol{j}_r
\end{pmatrix}
\end{equation}
The term ``generalised'' is used in order to account for the fact that the physical dimensions may differ in each term of the sum. In view of the generalised force vectors~\eqref{generalised force vectors} and current densities vectors~\eqref{generalised current density vectors}, the scalar internal power density~\eqref{power density int S} is recast as,
\begin{equation}\label{power density int S bis}
p^{\,\text{int}}_S = F^{\top} \cdot j = \sum_{i=0}^{n}\,F_i\,j_i \geqslant 0
\end{equation}
and the vectorial internal power density~\eqref{power density int S} is rewritten as,
\begin{equation}\label{power density int V bis}
p^{\,\text{int}}_V = \boldsymbol{F}^{\top} \cdot \boldsymbol{j} = \sum_{\alpha = 0}^{r}\,\boldsymbol{F}_{\alpha}\cdot\boldsymbol{j}_{\alpha} \geqslant 0
\end{equation}
where the transpose of the columns vectors $F$ and $\boldsymbol{F}$ are the line vectors $F^{\top}$ and $\boldsymbol{F}^{\top}$.

\section{Scalar Onsager reciprocal relations}
\label{Scalar Onsager reciprocal relations}
\noindent The generalised scalar force vector $F$ is the physical cause of the generalised scalar current density vector $j$. At equilibrium, the generalised scalar force vector $F$ vanishes and so does the generalised scalar current density vector $j$. Thus, according to relation~\eqref{power density int S bis}, there is no scalar internal power density at equilibrium, i.e. $p^{\,\text{int}}_S=0$. Out of equilibrium, in the neighbourhood of the equilibrium where the generalised scalar force vector $F$ is sufficiently small, the generalised scalar current density vector $j$ can be expanded to first-order in terms of the generalised scalar force vector $F$,~\cite{Brechet:2019} 
\begin{equation}\label{linear relation S}
j = L\,F
\end{equation}
where the scalar Onsager matrix $L \in M_{n+1}\left(\mathbb{R}\right)$ is written in scalar components as, 
\begin{equation}\label{matrice L S}
L = \begin{pmatrix}
L_{00} & L_{01} & \cdots & L_{0n} \\
L_{10} & L_{11} & \cdots & L_{1n} \\
\vdots & \vdots & \ddots & \vdots \\
L_{n0} & L_{n1} & \cdots & L_{nn} 
\end{pmatrix}
\end{equation}
On a mathematical level, in view of the scalar linear phenomenological relation~\eqref{linear relation S} in the neighbourhood of the equilibrium, the scalar internal power density~\eqref{power density int S bis} is a positive semi-definite quadratic form of the generalised scalar force vector $F$,
\begin{equation}\label{power density int S quad}
p^{\,\text{int}}_S = F^{\top}\,L\,F \geqslant 0
\end{equation}
Since the scalar internal power density $p^{\,\text{int}}_S = F^{\top}\,L\,F$ is a quadratic form, it is equal to its transpose,
\begin{equation}\label{power density S transpose}
F^{\top}\,L\,F = \left(F^{\top}\,L\,F\right)^{\top} = F^{\top}\,L^{\top}\,F
\end{equation}
The scalar Onsager matrix $L$ is written as the sum of a symmetric matrix $S$ and an antisymmetric matrix $A$,
\begin{equation}\label{scalar Onsager matrix split}
L = S + A \qquad\text{where}\qquad S = S^{\top} \qquad\text{and}\qquad A = -\,A^{\top}
\end{equation}
In view of the decomposition~\eqref{scalar Onsager matrix split}, the quadratic form~\eqref{power density S transpose} is recast as,
\begin{equation}\label{power density transpose S split}
\begin{split}
&F^{\top}\,S\,F + F^{\top}\,A\,F = \left(F^{\top}\,S\,F\right)^{\top} + \left(F^{\top}\,A\,F\right)^{\top}\\
&\phantom{F^{\top}\,S\,F + F^{\top}\,A\,F} = F^{\top}\,S^{\top}\,F + F^{\top}\,A^{\top}\,F = F^{\top}\,S\,F -\,F^{\top}\,A\,F
\end{split}
\end{equation}
which shows that the antisymmetric part of the positive semi-definite quadratic form does not contribute to the scalar internal power density $p^{\,\text{int}}_S$,
\begin{equation}\label{power density no antisymmetry S}
F^{\top}\,A\,F = -\,F^{\top}\,A^{\top}\,F = 0
\end{equation}
On a physical level, the generalised scalar force vector $F$ is the cause of the generalised scalar current force density vector $j$ in the scalar linear phenomenological relation~\eqref{linear relation S}. Since the generalised scalar force vector $F$ is dissipative, the generalised scalar current force density vector $j$ describes dissipative phenomena. According to relation~\eqref{power density no antisymmetry S}, the antisymmetric part of the scalar Onsager matrix $L$ does not account for the dissipation. Thus, it plays the role of a gauge freedom. Therefore, without lack of generality, we can ignore the antisymmetric part of the scalar Onsager matrix $L$,
\begin{equation}\label{power density no antisymmetry bis S}
A = -\,A^{\top} = 0
\end{equation}
which implies that the Onsager matrix is symmetric,
\begin{equation}\label{power density int S symmetric}
L = S = S^{\top} = L^{\top} 
\end{equation}
The symmetry of the scalar Onsager matrix~\eqref{power density int S quad} can be written in components which yields the scalar Onsager reciprocal relations,
\begin{equation}\label{Onsager relations S}
L_{ij} = L_{ji} \qquad\forall \quad i,j = 0,\ldots,n
\end{equation}
In view of the symmetry~\eqref{power density int S symmetric}, the symmetric scalar Onsager matrix $L$ can be written in terms of a square matrix $M \in M_{n+1}\left(\mathbb{R}\right)$ as,
\begin{equation}\label{power density int S ter bis}
L = M^{\top}M
\end{equation}
where the matrix $M$ is written in scalar components as, 
\begin{equation}\label{matrice M S}
M = \begin{pmatrix}
M_{00} & M_{01} & \cdots & M_{0n} \\
M_{10} & M_{11} & \cdots & M_{1n} \\
\vdots & \vdots & \ddots & \vdots \\
M_{n0} & M_{n1} & \cdots & M_{nn} 
\end{pmatrix}
\end{equation}
The symmetry of the scalar Onsager matrix~\eqref{power density int S ter bis} is then straightforward,
\begin{equation}\label{symmetry Onsager M S}
L^{\top} = \left(M^{\top}M\right)^{\top} = M^{\top}M = L
\end{equation}
The positive semi-definite quadratic form~\eqref{power density int S quad} can be recast as the square of the norm of the linear map $MF$,
\begin{equation}\label{power density int S pent}
p^{\,\text{int}}_S = F^{\top}\,L\,F = F^{\top}\,M^{\top}M\,F = \left(M\,F\right)^{\top}\cdot\left(M\,F\right) =
\Vert M\,F \Vert^2 \geqslant 0
\end{equation}
The symmetric scalar Onsager matrix~\eqref{power density int S ter bis} is written in components as,
\begin{equation}\label{L S components}
L_{ij} = \sum_{k=0}^{n}M^{\top}_{ik}\,M_{kj} = \sum_{k=0}^{n}M_{ki}\,M_{kj} = \sum_{k=0}^{n}M_{kj}\,M_{ki} = \sum_{k=0}^{n}M^{\top}_{jk}\,M_{ki} = L_{ji}
\end{equation}
Furthermore, the diagonal components of the scalar Onsager matrix~\eqref{L S components} are non-negative,
\begin{equation}\label{L S diagonal components}
L_{ii} = \sum_{k=0}^{n}M_{ki}^2 \geqslant 0
\end{equation}

\section{Scalar Onsager-Casimir reciprocal relations}
\label{Scalar Onsager-Casimir reciprocal relations}
\noindent The scalar Onsager-Casimir reciprocal relations are a generalisation of the scalar Onsager reciprocal relations~\eqref{Onsager relations S} accounting for the symmetry under time reversal. The local volume expansion rate density $\boldsymbol{\nabla}\cdot\boldsymbol{v}$ changes sign if the recordings of the thermodynamic processes are watched backwards in time and the chemical affinities $\mathcal{A}_1,\ldots,\mathcal{A}_n$ do not change sign. Thus, the time reversal $T$ of the generalised scalar force vector~\eqref{generalised force vectors} is written as,
\begin{equation}\label{time force column S}
T\,:\quad F = 
\begin{pmatrix} 
\boldsymbol{\nabla}\cdot\boldsymbol{v}\\
\mathcal{A}_1\\
\vdots\\
\mathcal{A}_n
\end{pmatrix}
\ \mapsto\ 
\begin{pmatrix} 
-\,\boldsymbol{\nabla}\cdot\boldsymbol{v}\\
\mathcal{A}_1\\
\vdots\\
\mathcal{A}_n
\end{pmatrix}
\end{equation}
The linear mapping corresponding to the time reversal of the generalised scalar force vector $F$ can thus be described by an idempotent diagonal matrix $\varepsilon \in M_{n+1}\left(\mathbb{R}\right)$,
\begin{equation}\label{epsilon matrix S}
\varepsilon = \begin{pmatrix}
\varepsilon_0 & 0 & \cdots & 0 \\
0 & \varepsilon_1 & \cdots & 0 \\
\vdots & \vdots & \ddots & \vdots\\
0 & 0 & \cdots & \varepsilon_n
\end{pmatrix}
=
\begin{pmatrix}
-\,1 & 0 & \cdots & 0 \\
0 & 1 & \cdots & 0 \\
\vdots & \vdots & \ddots & \vdots\\
0 & 0 & \cdots & 1
\end{pmatrix}
\end{equation}
which is a projector defined by,
\begin{equation}\label{projector S}
\varepsilon^2 = \mathbb{1}_{n+1} \qquad\text{and}\qquad \varepsilon^{-1} = \varepsilon^{\top} = \varepsilon
\end{equation}
where $\mathbb{1}_{n+1} \in M_{n+1}\left(\mathbb{R}\right)$ is the identity matrix. The scalar $\varepsilon_0 = -\,1$ is referred to as the odd parity, and the scalars $\varepsilon_1 = \dots = \varepsilon_n = 1$ as the even parity under time reversal in the literature~\cite{Pavelka:2018}. Since the scalar internal power density $p^{\,\text{int}}_S$ is a positive semi-definite quadratic form of the generalised scalar force vector $F$ accounting for the dissipation, it is invariant under time reversal $T$ : $F\ \mapsto\ \varepsilon\,F$. In view of relations~\eqref{symmetry Onsager M S},~\eqref{power density int S pent} and~\eqref{projector S}, this invariance is written as,
\begin{equation}\label{power density int S hex}
p^{\,\text{int}}_S = \Vert M\,\varepsilon\,F \Vert^2 = \left(M\,\varepsilon\,F\right)^{\top}\cdot\left(M\,\varepsilon\,F\right) = F^{\top}\,\varepsilon^{\top}\,M^{\top}M\,\varepsilon\,F = F^{\top}\,\varepsilon\,L\,\varepsilon\,F \geqslant 0
\end{equation}
Moreover, the scalar internal power density~\eqref{power density int S quad} is a scalar quantity that is equal to its transpose,
\begin{equation}\label{power density transpose S}
p^{\,\text{int}}_S = \left(F^{\top}\,L\,F\right)^{\top} = F^{\top}\,L^{\top}\,F \geqslant 0
\end{equation}
The identification of the scalar internal power densities~\eqref{power density int S hex} and~\eqref{power density transpose S}  yields the relation,
\begin{equation}\label{power density int S hep}
F^{\top}\,\varepsilon\,L\,\varepsilon\,F = F^{\top}\,L^{\top}\,F
\end{equation}
The identity of the quadratic forms~\eqref{power density int S hep} has to hold for any value of the generalised scalar force column vector $F$ and the generalised scalar force line vector $F^{\top}$, which implies that,
\begin{equation}\label{Onsager identity transpose S}
\varepsilon\,L\,\varepsilon = L^{\top} 
\end{equation}
In view of the scalar Onsager matrix~\eqref{matrice L S} and the scalar time reversal matrix~\eqref{epsilon matrix S}, the scalar components of relation~\eqref{Onsager identity transpose S} yield the scalar Onsager-Casimir reciprocal relations,~\cite{Jou:2010}
\begin{equation}\label{Onsager-Casimir relations S}
\varepsilon_{i}\,\varepsilon_{j}\,L_{ij} = L_{ji} \qquad\forall \quad i,j = 0,\ldots,n
\end{equation}

\section{Vectorial Onsager reciprocal relations}
\label{Vectorial Onsager reciprocal relations}
\noindent The generalised vectorial force vector $\boldsymbol{F}$ is the physical cause of the generalised vectorial current density vector $\boldsymbol{j}$. At equilibrium, the generalised vectorial force vector $\boldsymbol{F}$ vanishes and so does the generalised vectorial current density vector $\boldsymbol{j}$. Thus, according to relation~\eqref{power density int V bis}, there is no vectorial internal power density at equilibrium, i.e. $p^{\,\text{int}}_V=0$. Out of equilibrium, in the neighbourhood of the equilibrium where the generalised vectorial force vector $\boldsymbol{F}$ is sufficiently small, the generalised vectorial current density vector $\boldsymbol{j}$ can be expanded to first-order in terms of the generalised vectorial force vector $\boldsymbol{F}$,
\begin{equation}\label{linear relation V}
\boldsymbol{j} = \boldsymbol{L}\,\boldsymbol{F}
\end{equation}
where the vectorial Onsager matrix $\boldsymbol{L} \in M_{r+1}\left(\mathbb{R}^3\right)$ is written in tensorial components as,
\begin{equation}\label{matrice L V}
\boldsymbol{L} = \begin{pmatrix}
\boldsymbol{L}_{00} & \boldsymbol{L}_{01} & \cdots & \boldsymbol{L}_{0r} \\
\boldsymbol{L}_{10} & \boldsymbol{L}_{11} & \cdots & \boldsymbol{L}_{1r} \\
\vdots & \vdots & \ddots & \vdots \\
\boldsymbol{L}_{r0} & \boldsymbol{L}_{r1} & \cdots & \boldsymbol{L}_{rr} 
\end{pmatrix}
\end{equation}
On a mathematical level, in view of the vectorial linear phenomenological relation~\eqref{linear relation V} in the neighbourhood of the equilibrium, the vectorial internal power density~\eqref{power density int V bis} is a positive semi-definite quadratic form of the generalised vectorial force vector $\boldsymbol{F}$,
\begin{equation}\label{power density int V quad}
p^{\,\text{int}}_V = \boldsymbol{F}^{\top}\boldsymbol{L}\,\boldsymbol{F} \geqslant 0
\end{equation}
Since the vectorial internal power density $p^{\,\text{int}}_V = \boldsymbol{F}^{\top}\,\boldsymbol{L}\,\boldsymbol{F}$ is a quadratic form, it is equal to its transpose,
\begin{equation}\label{power density V transpose}
\boldsymbol{F}^{\top}\,\boldsymbol{L}\,\boldsymbol{F} = \left(\boldsymbol{F}^{\top}\,\boldsymbol{L}\,\boldsymbol{F}\right)^{\top} = \boldsymbol{F}^{\top}\,\boldsymbol{L}^{\top}\,\boldsymbol{F}
\end{equation}
The vectorial Onsager matrix $\boldsymbol{L}$ is written as the sum of a symmetric matrix $\boldsymbol{S}$ and an antisymmetric matrix $\boldsymbol{A}$,
\begin{equation}\label{vectorial Onsager matrix split}
\boldsymbol{L} = \boldsymbol{S} + \boldsymbol{A} \qquad\text{where}\qquad \boldsymbol{S} = \boldsymbol{S}^{\top} \qquad\text{and}\qquad \boldsymbol{A} = -\,\boldsymbol{A}^{\top}
\end{equation}
In view of the decomposition~\eqref{vectorial Onsager matrix split}, the quadratic form~\eqref{power density V transpose} is recast as,
\begin{equation}\label{power density transpose V split}
\begin{split}
&\boldsymbol{F}^{\top}\,\boldsymbol{S}\,\boldsymbol{F} + \boldsymbol{F}^{\top}\,\boldsymbol{A}\,\boldsymbol{F} = \left(\boldsymbol{F}^{\top}\,\boldsymbol{S}\,\boldsymbol{F}\right)^{\top} + \left(\boldsymbol{F}^{\top}\,\boldsymbol{A}\,\boldsymbol{F}\right)^{\top}\\
&\phantom{\boldsymbol{F}^{\top}\,\boldsymbol{S}\,\boldsymbol{F} + \boldsymbol{F}^{\top}\,\boldsymbol{A}\,\boldsymbol{F}} =  \boldsymbol{F}^{\top}\,\boldsymbol{S}^{\top}\,\boldsymbol{F} + \boldsymbol{F}^{\top}\,\boldsymbol{A}^{\top}\,\boldsymbol{F} = \boldsymbol{F}^{\top}\,\boldsymbol{S}\,\boldsymbol{F} -\,\boldsymbol{F}^{\top}\,\boldsymbol{A}\,\boldsymbol{F}
\end{split}
\end{equation}
which shows that the antisymmetric part of the positive semi-definite quadratic form does not contribute to the vectorial internal power density $p^{\,\text{int}}_V$,
\begin{equation}\label{power density no antisymmetry V}
\boldsymbol{F}^{\top}\,\boldsymbol{A}\,\boldsymbol{F} = -\,\boldsymbol{F}^{\top}\,\boldsymbol{A}^{\top}\,\boldsymbol{F} = 0
\end{equation}
On a physical level, the generalised vectorial force vector $\boldsymbol{F}$ is the cause of the generalised vectorial current force density vector $\boldsymbol{j}$ in the vectorial linear phenomenological relation~\eqref{linear relation V}. Since the generalised vectorial force vector $\boldsymbol{F}$ is dissipative, the generalised vectorial current force density vector $\boldsymbol{j}$ describes dissipative phenomena. According to relation~\eqref{power density no antisymmetry V}, the antisymmetric part of the vectorial Onsager matrix $\boldsymbol{L}$ does not account for the dissipation. Thus, it plays the role of a gauge freedom. Therefore, without lack of generality, we can ignore the antisymmetric part of the vectorial Onsager matrix $\boldsymbol{L}$,
\begin{equation}\label{power density no antisymmetry bis V}
\boldsymbol{A} = -\,\boldsymbol{A}^{\top} = \boldsymbol{0}
\end{equation}
which implies that the vectorial Onsager matrix is symmetric,
\begin{equation}\label{power density int V symmetric}
\boldsymbol{L} = \boldsymbol{S} = \boldsymbol{S}^{\top} = \boldsymbol{L}^{\top} 
\end{equation}
The symmetry of the vectorial Onsager matrix~\eqref{power density int V quad} can be written in components which yields the vectorial Onsager reciprocal relations,
\begin{equation}\label{Onsager relations V}
\boldsymbol{L}_{\alpha\beta} = \boldsymbol{L}_{\beta\alpha} \qquad\forall \quad \alpha,\beta = 0,\ldots,r
\end{equation}
In view of the symmetry~\eqref{power density int V symmetric}, the symmetric vectorial Onsager matrix $\boldsymbol{L}$ can be written in terms of a square matrix $\boldsymbol{M} \in M_{r+1}\left(\mathbb{R}^3\right)$ as,
\begin{equation}\label{power density int V ter bis}
\boldsymbol{L} = \boldsymbol{M}^{\top}\boldsymbol{M}
\end{equation}
where the matrix $\boldsymbol{M}$ is written in tensorial components as, 
\begin{equation}\label{matrice M V}
\boldsymbol{M} = \begin{pmatrix}
\boldsymbol{M}_{00} & \boldsymbol{M}_{01} & \cdots & \boldsymbol{M}_{0r} \\
\boldsymbol{M}_{10} & \boldsymbol{M}_{11} & \cdots & \boldsymbol{M}_{1r} \\
\vdots & \vdots & \ddots & \vdots \\
\boldsymbol{M}_{r0} & \boldsymbol{M}_{r1} & \cdots & \boldsymbol{M}_{rr} 
\end{pmatrix}
\end{equation}
The symmetry of the vectorial Onsager matrix~\eqref{power density int V ter bis} is then straightforward,
\begin{equation}\label{symmetry Onsager M V}
\boldsymbol{L}^{\top} = \left(\boldsymbol{M}^{\top}\boldsymbol{M}\right)^{\top} = \boldsymbol{M}^{\top}\boldsymbol{M} = \boldsymbol{L}
\end{equation}
The positive semi-definite quadratic form~\eqref{power density int V quad} can be recast as the square of the norm of the linear map $\boldsymbol{M}\,\boldsymbol{F}$,
\begin{equation}\label{power density int V pent}
p^{\,\text{int}}_V = \boldsymbol{F}^{\top}\,L\,\boldsymbol{F} = \boldsymbol{F}^{\top}\,\boldsymbol{M}^{\top}\boldsymbol{M}\,\boldsymbol{F} = \left(\boldsymbol{M}\,\boldsymbol{F}\right)^{\top}\cdot\left(\boldsymbol{M}\,\boldsymbol{F}\right) =
\Vert \boldsymbol{M}\,\boldsymbol{F} \Vert^2 \geqslant 0
\end{equation}
The symmetric vectorial Onsager matrix~\eqref{power density int V ter bis} is written in components as,
\begin{equation}\label{L V components}
\boldsymbol{L}_{\alpha\beta} = \sum_{\gamma=0}^{r}\boldsymbol{M}^{\top}_{\alpha\gamma}\,\boldsymbol{M}_{\gamma\beta} = \sum_{\gamma=0}^{r}\boldsymbol{M}_{\gamma\alpha}\,\boldsymbol{M}_{\gamma\beta} = \sum_{\gamma=0}^{r}\boldsymbol{M}_{\gamma\beta}\,\boldsymbol{M}_{\gamma\alpha} = \sum_{\gamma=0}^{r}\boldsymbol{M}^{\top}_{\beta\gamma}\,\boldsymbol{M}_{\gamma\alpha} = \boldsymbol{L}_{\beta\alpha}
\end{equation}
The eigenspaces of these spatial matrices are determined by the spatial symmetries of the thermodynamic system. In particular, the presence of a magnetic field $\boldsymbol{B}$ leads to the same spatial symmetry breaking for different irreversible processes such as the Hall effect, the Nernst effect or the Righi-Leduc effect. Therefore, these eigenspaces are the same for every spatial matrix $\boldsymbol{M}_{\alpha\beta}$ where $\alpha,\beta \in \{0,..,r\}$, which implies that these matrices commute. Furthermore, the diagonal components of the vectorial Onsager matrix~\eqref{L V components} are non-negative,
\begin{equation}\label{L V diagonal components}
\boldsymbol{L}_{\alpha\alpha} = \sum_{\gamma=0}^{r}\boldsymbol{M}_{\gamma\alpha}^2 \geqslant 0
\end{equation}

\section{Vectorial Onsager-Casimir reciprocal relations}
\label{Vectorial Onsager-Casimir reciprocal relations}
\noindent The vectorial Onsager-Casimir reciprocal relations are a generalisation of the vectorial Onsager reciprocal relations~\eqref{Onsager relations V} accounting for the symmetries of the vectorial linear relations~\eqref{linear relation V} under time reversal. The temperature gradient $\boldsymbol{\nabla}\,T$ and the electrochemical gradient $\boldsymbol{\nabla}\,\bar{\mu}_1,\ldots,\boldsymbol{\nabla}\,\bar{\mu}_r$ of the chemical substances that do not change sign at a specific time if the recordings of the thermodynamic processes are watched backwards in time. Thus, the time reversal $T$ of the generalised vectorial force vector~\eqref{generalised force vectors} is written as,
\begin{equation}\label{time force column V}
T\,:\quad \boldsymbol{F} = 
\begin{pmatrix} 
-\,\boldsymbol{\nabla}\,T\\
-\,\boldsymbol{\nabla}\,\bar{\mu}_1\\
\vdots\\
-\,\boldsymbol{\nabla}\,\bar{\mu}_r
\end{pmatrix}
\ \mapsto\ 
\begin{pmatrix} 
-\,\boldsymbol{\nabla}\,T\\
-\,\boldsymbol{\nabla}\,\bar{\mu}_1\\
\vdots\\
-\,\boldsymbol{\nabla}\,\bar{\mu}_r
\end{pmatrix}
\end{equation}
The time symmetry of vectorial quantities in thermodynamic systems is discussed notably by Pavelka et al~\cite{Pavelka:2018}. The linear mapping corresponding to the time reversal of the generalised vectorial force vector $\boldsymbol{F}$ can thus be described by an idempotent diagonal matrix $\varepsilon \in M_{r+1}\left(\mathbb{R}^3\right)$,
\begin{equation}\label{epsilon matrix V}
\boldsymbol{\varepsilon} = \begin{pmatrix}
\varepsilon_0\,\mathbb{1}_3 & 0 & \cdots & 0 \\
0 & \varepsilon_1\,\mathbb{1}_3 & \cdots & 0 \\
\vdots & \vdots & \ddots & \vdots\\
0 & 0 & \cdots & \varepsilon_r\,\mathbb{1}_3
\end{pmatrix}
=
\begin{pmatrix}
\mathbb{1}_3 & 0 & \cdots & 0 \\
0 & \mathbb{1}_3 & \cdots & 0 \\
\vdots & \vdots & \ddots & \vdots\\
0 & 0 & \cdots & \,\mathbb{1}_3
\end{pmatrix}
\end{equation}
which is a projector defined by,
\begin{equation}\label{projector V}
\boldsymbol{\varepsilon}^2 = \mathbb{1}_{3r+3} \qquad\text{and}\qquad \boldsymbol{\varepsilon}^{-1} = \boldsymbol{\varepsilon}^{\top} = \boldsymbol{\varepsilon}
\end{equation}
where $\mathbb{1}_{3}\in M_{3}\left(\mathbb{R}\right)$ and $\mathbb{1}_{3r+3} \in M_{r+1}\left(\mathbb{R}^3\right)$ are the identity matrices and $\varepsilon_0, \ldots, \varepsilon_r$ are the parities under time reversal~\cite{Pavelka:2018}. Since the vectorial internal power density $p^{\,\text{int}}_V$ is a positive semi-definite quadratic form of the generalised vectorial force vector $\boldsymbol{F}$ accounting for the dissipation, it is invariant under time reversal $T$ : $\boldsymbol{F}\ \mapsto\ \boldsymbol{\varepsilon}\,\boldsymbol{F}$. In view of relations~\eqref{symmetry Onsager M V},~\eqref{power density int V pent} and~\eqref{projector V}, this invariance is written as,
\begin{equation}\label{power density int V hex}
p^{\,\text{int}}_V = \Vert \boldsymbol{M}\,\boldsymbol{\varepsilon}\,\boldsymbol{F} \Vert^2 = \left(\boldsymbol{M}\,\boldsymbol{\varepsilon}\,\boldsymbol{F}\right)^{\top}\cdot\left(\boldsymbol{M}\,\boldsymbol{\varepsilon}\,\boldsymbol{F}\right) = \boldsymbol{F}^{\top}\,\boldsymbol{\varepsilon}^{\top}\,\boldsymbol{M}^{\top}\boldsymbol{M}\,\boldsymbol{\varepsilon}\,\boldsymbol{F} = \boldsymbol{F}^{\top}\,\boldsymbol{\varepsilon}\,\boldsymbol{L}\,\boldsymbol{\varepsilon}\,\boldsymbol{F} \geqslant 0
\end{equation}
Moreover, the vectorial internal power density~\eqref{power density int V quad} is a scalar quantity that is equal to its transpose,
\begin{equation}\label{power density transpose V}
p^{\,\text{int}}_V = \left(\boldsymbol{F}^{\top}\,\boldsymbol{L}\,\boldsymbol{F}\right)^{\top} = \boldsymbol{F}^{\top}\,\boldsymbol{L}^{\top}\,\boldsymbol{F} \geqslant 0
\end{equation}
The identification of the vectorial internal power densities~\eqref{power density int V hex} and~\eqref{power density transpose V}  yields the relation,
\begin{equation}\label{power density int V hep}
\boldsymbol{F}^{\top}\,\boldsymbol{\varepsilon}\,\boldsymbol{L}\,\boldsymbol{\varepsilon}\,\boldsymbol{F} = \boldsymbol{F}^{\top}\,\boldsymbol{L}^{\top}\,\boldsymbol{F}
\end{equation}
The identity of the quadratic forms~\eqref{power density int V hep} has to hold for any value of the generalised vectorial force column vector $\boldsymbol{F}$ and the generalised vectorial force line vector $\boldsymbol{F}^{\top}$, which implies that,
\begin{equation}\label{Onsager identity transpose V}
\boldsymbol{\varepsilon}\,\boldsymbol{L}\,\boldsymbol{\varepsilon} = \boldsymbol{L}^{\top} 
\end{equation}
In view of the vectorial Onsager matrix~\eqref{matrice L V} and the vectorial time reversal matrix~\eqref{epsilon matrix V}, the tensorial components of relation~\eqref{Onsager identity transpose V} yield the vectorial Onsager-Casimir reciprocal relations,~\cite{Jou:2010}
\begin{equation}\label{Onsager-Casimir relations V}
\varepsilon_{\alpha}\,\varepsilon_{\beta}\,\boldsymbol{L}_{\alpha\beta} = \boldsymbol{L}_{\beta\alpha} \qquad\forall \quad \alpha,\beta = 0,\ldots,r
\end{equation}
In the presence of an external magnetic field $\boldsymbol{B}$ and an angular velocity vector field $\boldsymbol{\omega}$, due to the rotation of the thermodynamic system, the field dependance of the components of the vectorial Onsager matrix has to be specified in the vectorial Onsager-Casimir reciprocal relations. Under time reversal $T$, the pseudovectors magnetic field $\boldsymbol{B}$ and angular velocity $\boldsymbol{\omega}$ change sign,~\cite{Jou:2010}
\begin{equation}\label{magnetic projection}
T\,:\quad\,\boldsymbol{B}\ \mapsto\ -\,\boldsymbol{B} \qquad\text{and}\qquad \boldsymbol{\omega}\ \mapsto\ -\,\boldsymbol{\omega}
\end{equation}
Thus, the vectorial Onsager-Casimir reciprocal relations~\eqref{Onsager-Casimir relations V} become,
\begin{equation}\label{Onsager-Casimir relations bis V}
\boldsymbol{L}_{\beta\alpha}\left(\boldsymbol{B},\boldsymbol{\omega}\right)  = \varepsilon_{\alpha}\,\varepsilon_{\beta}\,\boldsymbol{L}_{\alpha\beta}\left(-\,\boldsymbol{B},-\,\boldsymbol{\omega}\right) 
\end{equation}
These reciprocal relations can be recast in a more traditional way as it is usually done in the literature notably by as Pavelka et al.~\cite{Pavelka:2018},
\begin{align}\label{Onsager-Casimir relations VI}
&\boldsymbol{L}_{\beta\alpha}\left(\boldsymbol{B},\boldsymbol{\omega}\right)  = \boldsymbol{L}_{\alpha\beta}\left(-\,\boldsymbol{B},-\,\boldsymbol{\omega}\right) \phantom{-\,}\qquad\text{for}\qquad \varepsilon_{\alpha} = \varepsilon_{\beta} = \pm 1\\
&\boldsymbol{L}_{\beta\alpha}\left(\boldsymbol{B},\boldsymbol{\omega}\right)  = -\,\boldsymbol{L}_{\alpha\beta}\left(-\,\boldsymbol{B},-\,\boldsymbol{\omega}\right) \qquad\text{for}\qquad \varepsilon_{\alpha} = -\,\varepsilon_{\beta} = \pm 1
\end{align}
%


\section{Conclusion}
\label{Conclusion}

\noindent The local expression of the second law of thermodynamics for a continuous medium requires the internal power density, accounting for the dissipation due to the irreversible processes, to be expressed as the sum of two positive definite quadratic forms of the generalised forces in the neighbourhood of the equilibrium : the scalar internal power density $p_S^{\,\text{int}}$ describing the chemistry and the vectorial density $p_V^{\,\text{int}}$ describing the transport. The scalar internal power density is a positive semi-definite quadratic form of the generalised scalar force vector $F$,
\begin{equation}\label{power density int S end}
p^{\,\text{int}}_S = F^{\top}\,L\,F \geqslant 0
\end{equation}
The scalar Onsager matrix $L$ is written as the sum of a symmetric matrix $S$ and an antisymmetric matrix $A$,
\begin{equation}\label{scalar Onsager matrix split end}
L = S + A \qquad\text{where}\qquad S = S^{\top} \qquad\text{and}\qquad A = -\,A^{\top}
\end{equation}
Since the antisymmetric part of the scalar Onsager matrix $L$ does not account for the dissipation, it plays the role of a gauge freedom, which implies that the scalar Onsager matrix is symmetric,
\begin{equation}\label{power density int S symmetric end}
L = S = S^{\top} = L^{\top} 
\end{equation}
The scalar components of the scalar Onsager matrix~\eqref{power density int S symmetric end} yields the scalar Onsager reciprocal relations,
\begin{equation}\label{Onsager relations S end}
L_{ij} = L_{ji} \qquad\forall \quad i,j = 0,\ldots,n
\end{equation}
Similarly, the vectorial internal power density is a positive semi-definite quadratic form of the generalised vectorial force vector $\boldsymbol{F}$,
\begin{equation}\label{power density int V end}
p^{\,\text{int}}_V = \boldsymbol{F}^{\top}\,\boldsymbol{L}\,\boldsymbol{F} \geqslant 0
\end{equation}
The vectorial Onsager matrix $\boldsymbol{L}$ is written as the sum of a symmetric matrix $\boldsymbol{S}$ and an antisymmetric matrix $\boldsymbol{A}$,
\begin{equation}\label{vectorial Onsager matrix split end}
\boldsymbol{L} = \boldsymbol{S} + \boldsymbol{A} \qquad\text{where}\qquad \boldsymbol{S} = \boldsymbol{S}^{\top} \qquad\text{and}\qquad \boldsymbol{A} = -\,\boldsymbol{A}^{\top}
\end{equation}
Since the antisymmetric part of the vectorial Onsager matrix $\boldsymbol{L}$ does not account for the dissipation, it plays the role of a gauge freedom, which implies that the vectorial Onsager matrix is symmetric,
\begin{equation}\label{power density int V symmetric end}
\boldsymbol{L} = \boldsymbol{S} = \boldsymbol{S}^{\top} = \boldsymbol{L}^{\top} 
\end{equation}
The tensorial components of the vectorial Onsager matrix~\eqref{power density int V symmetric end} yields the vectorial Onsager reciprocal relations,
\begin{equation}\label{Onsager relations V end}
\boldsymbol{L}_{\alpha\beta} = \boldsymbol{L}_{\beta\alpha} \qquad\forall \quad \alpha,\beta = 0,\ldots,r
\end{equation}
The scalar and vectorial Casimir-Onsager reciprocal relations are the generalisation the scalar and vectorial Onsager reciprocal relations. They are derived by taking into account the symmetries of the generalised force vectors under time reversal. The action of the time reversal $T$ on the scalar generalised force column vector $F$ is written in terms of the projection matrix $\varepsilon$. The scalar internal power density $p^{\,\text{int}}_S$ is a positive semi-definite quadratic form of the generalised scalar force vector $F$ that is invariant under time reversal $T$ : $F\ \mapsto\ \varepsilon\,F$. Moreover, it is equal to its transpose,
\begin{equation}\label{power density int S hep end}
p^{\,\text{int}}_S = F^{\top}\,\varepsilon\,L\,\varepsilon\,F = F^{\top}\,L^{\top}\,F \geqslant 0
\end{equation}
which implies that,
\begin{equation}\label{power density int S bis end}
\varepsilon\,L\,\varepsilon = L^{\top}
\end{equation}
and yields the scalar Onsager-Casimir reciprocal relations,
\begin{equation}\label{Onsager-Casimir relations S end}
\varepsilon_{i}\,\varepsilon_{j}\,L_{ij} = L_{ji} \qquad\forall \quad i,j = 0,\ldots,n
\end{equation}d
Similarly, the action of the time reversal $T$ on the vectorial generalised force column vector $\boldsymbol{F}$ is written in terms of the projection matrix $\boldsymbol{\varepsilon}$. The vectorial internal power density $p^{\,\text{int}}_V$ is a positive semi- definite quadratic form of the generalised vectorial vector $\boldsymbol{F}$ that is invariant under time reversal $T$ : $\boldsymbol{F}\ \mapsto\ \boldsymbol{\varepsilon}\,\boldsymbol{F}$. Moreover, it is equal to its transpose,
\begin{equation}\label{power density int V hep end}
p^{\,\text{int}}_S = \boldsymbol{F}^{\top}\,\boldsymbol{\varepsilon}\,\boldsymbol{L}\,\boldsymbol{\varepsilon}\,\boldsymbol{F} = \boldsymbol{F}^{\top}\,\boldsymbol{L}^{\top}\,\boldsymbol{F} \geqslant 0
\end{equation}
which implies that,
\begin{equation}\label{power density int V bis end}
\boldsymbol{\varepsilon}\,\boldsymbol{L}\,\boldsymbol{\varepsilon} = \boldsymbol{L}^{\top}
\end{equation}
and yields the vectorial Onsager-Casimir reciprocal relations,
\begin{equation}\label{Onsager-Casimir relations V fine}
\varepsilon_{\alpha}\,\varepsilon_{\beta}\,\boldsymbol{L}_{\alpha\beta} = \boldsymbol{L}_{\beta\alpha} \qquad\forall \quad \alpha,\beta = 0,\ldots,r
\end{equation}
In the presence of an external magnetic field $\boldsymbol{B}$ and an angular velocity vector field $\boldsymbol{\omega}$ the vectorial Onsager-Casimir reciprocal relations~\eqref{Onsager-Casimir relations V fine} become,
\begin{align}\label{Onsager-Casimir relations VI fine}
&\boldsymbol{L}_{\beta\alpha}\left(\boldsymbol{B},\boldsymbol{\omega}\right)  = \boldsymbol{L}_{\alpha\beta}\left(-\,\boldsymbol{B},-\,\boldsymbol{\omega}\right) \phantom{-\,}\qquad\text{for}\qquad \varepsilon_{\alpha} = \varepsilon_{\beta} = \pm 1\\
&\boldsymbol{L}_{\beta\alpha}\left(\boldsymbol{B},\boldsymbol{\omega}\right)  = -\,\boldsymbol{L}_{\alpha\beta}\left(-\,\boldsymbol{B},-\,\boldsymbol{\omega}\right) \qquad\text{for}\qquad \varepsilon_{\alpha} = -\,\varepsilon_{\beta} = \pm 1
\end{align}
The phenomenological derivation of the Onsager and Onsager-Casimir reciprocal relations within the framework of the thermodynamics of irreversible processes presented in this article is clearly much simpler and more straightforward then the relations established by Onsager~\cite{Onsager:1931a,Onsager:1931b} and Casimir~\cite{Casimir:1945} within the framework of statistical physics.


\section*{Acknowledgement}
\noindent The author would like to thank Jean-Philippe Ansermet and Fran\c{c}ois Reuse for insightful discussions as well as Christian Gruber for his inspirational work.


\bibliography{references}
\bibliographystyle{plainnat} 

\end{document}